# OpenMosix approach to build scalable HPC farms with an easy management infrastructure

R. Esposito, P. Mastroserio, G. Tortone
*INFN, Napoli, I-80126, Italy*

F. M. Taurino
*INFM, Napoli, I-80126, Italy*

The OpenMosix approach is a good solution to build powerful and scalable computing farms. Furthermore an easy management infrastructure is implemented using diskless nodes and network boot procedures. In HENP environment, the choice of OpenMosix has been proven to be an optimal solution to give a general performance boost on implemented systems thanks to its load balancing and process migration features. In this poster we give an overview of the activities, carried out by our computing center, concerning the installation, management, monitoring, and usage of HPC Linux clusters running OpenMosix.

## 1. INTRODUCTION

This document is a short report of the activities carried out by the computing centre of INFN, INFM and Dept. of Physics at University of Naples "Federico II", concerning the installation, management and usage of HPC Linux clusters running OpenMosix [1].

These activities started on small clusters, running a local copy of the operating system and evolved through the years to a configuration of computing farms based on a diskless architecture, simplifying installation and management tasks.

Thanks to high performances offered by low cost systems (common personal computers) and availability of operating systems like Linux, we were able to implement HPC clusters using standard PCs, connected to an efficient LAN.

This solution offers several advantages:
- lower costs than monolithic systems;
- common technologies and components available trough the retailers (COTS, Components Off The Shelf);
- high scalability.

## 2. OPENMOSIX

OpenMosix is a software package that was specifically designed to enhance the Linux kernel with cluster computing capabilities. One of the most interesting features of OpenMosix is the optimized management of computing resources. The migration of the workload is allowed from a node to another, in a pre-emptive and transparent way: users view a Single System Image Cluster. This accomplished by an efficient load balancing algorithm which also prevents the trashing of applications in case of memory-swapping. The adaptive resource sharing algorithms solve the problem of dynamic balancing of the workload on the Farm CPUs. They use, when necessary, the PPM (Pre-emptive Process Migration) module for load reallocation among the Farm CPUs. The migration takes place automatically on the basis of the conditions reported by the optimization of resource sharing, load balancing algorithms and, in particular, the individual load of the single nodes and the network speed.

Each process has a *Unique Home-Node* (UHN) where it was created. Normally this is the node to which the user has logged-in. The single system image model of OpenMosix is a CC (cache coherent) cluster, in which every process seems to run at its UHN, and all the processes of a user's session share the execution environment of the UHN. Processes that migrate to other (remote) nodes use local (in the remote node) resources whenever possible, but interact with the user's environment through the UHN. For example, assume that a user launches several processes, some of which migrate away from the UHN. If the user executes "*ps*", it will report the status of all the processes, including processes that are executing on remote nodes. If one of the migrated processes reads the current time by invoking *gettimeofday()*, it will get the current time at the UHN [2].

## 3. LINUX HPC FARMS

### 3.1. Management issues

Configuring a farm presents some difficulties due mainly to the high number of hardware elements to manage and control.

Farm setup can be a time-consuming and difficult task and the probability to make mistakes during the configuration process gets higher when the system manager has to install a large number of nodes.

Moreover, installing a new software package, upgrading a library, or even removing a program, become heavy tasks when they have to be replicated on every computing node.

A farm, at last, can be a complex "instrument" to use, since executing parallel jobs often requires an "a priori" knowledge about hosts status and a manual resources allocation.

The software architecture of the clusters we have implemented tries to solve such problems, offering an easier way to install, manage and share resources.

## 3.2. Proposed infrastructure

The model we have chosen is based on a Master Node that exports the operating system to the diskless nodes; a private network, commonly based on a Fast Ethernet switch, is used by slave nodes to remotely execute the bootstrap procedure and to mount the root filesystem from the boot server. A second private network based on Gigabit, Myrinet or QSNet is used for process migration and for high speed communications, such as in MPI applications.

Each machine, except the master node, has no system disk. This solution offers several advantages:
- installing and upgrading software on the farm gets very simple as every modification involves only the master node;
- increased reliability, because there are less mechanical components;
- reduced cost, because only the master node mounts a local hard disk;

This type of setup significantly reduces the management time for "slave" nodes.

## 4. IMPLEMENTATION

Network boot of diskless nodes can be achieved in different ways. A first solution is EtherBoot [3], an opensource tool that allows to create a ROM containing a startup code, depending on the Ethernet card installed on the diskless machines. This code can be loaded on a boot device, such as a floppy, a hard disk, or a network adapter EERPOM.

A second solution could be PXELinux [4], an opensource boot loader that allows machines to download and execute their kernel via network, using a PXE (Pre eXecution Environment) compliant ethernet cards.

We can shortly explain the boot sequence of a generic diskless node:

1. startup code, generated by EtherBoot, is loaded and executed at boot time. This step could be substituted by a PXE boot request;
2. the node sends on the network (via bootp or dhcp) a request to obtain an IP address from the server;
3. once obtained the IP address, the node downloads its OpenMosix kernel from the server, using TFTP protocol;
4. the OpenMosix kernel is loaded and the node begins the real operating system boot;
5. kernel, previously compiled with the "root filesystem over NFS" option, mounts its root filesystem via NFS from the server;
6. once mounted the root filesystem, the node completes its boot sequence initialising system services and, eventually, mounting local disks

Since every diskless node has to mount its root filesystem via NFS, master node should export an independent root filesystem image to each client.

Exporting the same root filesystem to every client is not convenient for obvious reasons, for example configurations and log files have to be different for each machine.

To solve this problem we chose to install ClusterNFS, an enhanced version of standard NFS server [5].

ClusterNFS is a an opensource tool which allows every machine in the cluster, (included the master node), to share the same root filesystem, paying attention to some syntax rules:
- all files are shared by default;
- an *xxx* file, common to every client but not to the server, has to be named *xxx$$CLIENT$$*;
- an *xxx* file, for a specific client has to be named *xxx$$HOST=hostname$$*
  or *xxx$$IP=111.222.333.444*, where *hostname* and *111.222.333.444* are the real hostname or the ip address of that node.

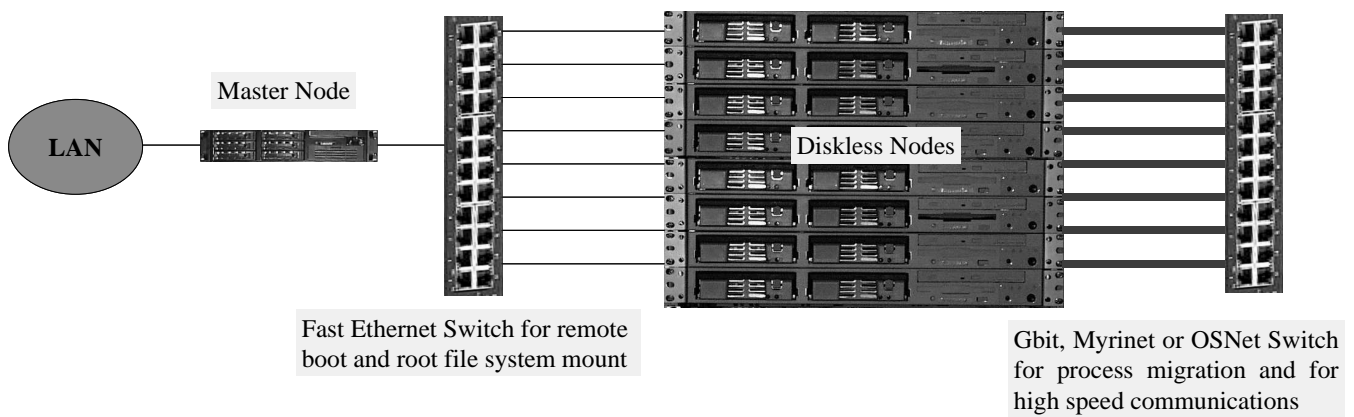

Figure 1: Farm layout

**TUDT002**

## 5. CONCLUSIONS

The setup procedure shown in this poster has been successfully replicated several times at INFN and INFM, in Naples, on Linux farms using many types of hardware and running different applications, such as:

- gravitational waves analysis (VIRGO project)
- cosmic rays analysis (ARGO project)
- MonteCarlo simulations

For our purposes, the most noticeable features of OpenMosix are its load balancing and process migration algorithms, which implies that users don't need to have knowledge of the current state of computing nodes. Parallel application can be executed by forking many processes, just like in an SMP machine, as OpenMosix continuously attempts to optimise the resource allocation.

The "OpenMosix + EtherBoot/PXELinux + ClusterNFS" approach is a good solution to build powerful computing farms with minimal installation and management effort. In our environment, this solution gave a general performance and productivity boost on implemented systems.

**TUDT002**